\documentstyle[12pt]{article}
\makeatletter
\@addtoreset{equation}{section}
\makeatother

\def\ee{\end{equation}}

\title{Nt-yet}
\begin{document}

\begin{flushright}
\vspace{1mm}
FIAN/TD/
 june {2002}\\
\end{flushright}

\vspace{1cm}

\begin{center}
{\large\bf Coincidence of large numbers, exact value of cosmological parameters and their analytical
representation }
\vglue 0.6  true cm
\vskip1cm
{\bf V.E.~Shemi-zadeh\footnotemark[1]} \footnotetext[1]{Internet adress: www.shemizadeh.narod.ru}
\vglue 0.3  true cm

I.E.Tamm Department of Theoretical Physics, Lebedev Physical Institute,\\
Leninsky prospect 53, 119991, Moscow, Russia
\vskip2cm
\end{center}

\begin{abstract}
A new approach to the phenomenon of large numbers coincidence leads to unexpected 
results. No matter how strange it might sound, the exact value of cosmological 
parameters and their analytical expression through fundamental constants have been 
founded. The basis for obtaining these unusual results is the equality of the 
fundamental Large Number to the exponent of the inverse value of the fine structure 
constant.
\end{abstract}
$$ $$
{\bf 1. INTRODUCTION}
{\smallskip}

It should be mentioned at the very beginning, that the work under consideration 
is unusual in form, content and results. Without making up any theoretical 
constructions, this research only compares and analyses figure value of observed data: 
fundamental constants and cosmological parameters. The work is the generalization 
and continuation of the problem of large numbers coincidence. From the point of view 
of microphysics and cosmology, the research is simple and illustrative, and could even 
be carried out by a student, at least by an interested one, who has some idea about 
Plank numbers, cosmological parameters and natural logarithms. Popularizing  the 
methodological essence of the work, one can say that the research concerns physical 
numerology, and certain manipulations of  physical numbers. In this respect, we found 
it appropriate to touch upon one number phenomenon which we called the Piazzi 
Smyth Effect and which is directly connected with the methodology of the present 
work.

Charles Piazzi Smyth is a well-known English astronomer, who, after visiting 
Egypt in the middle of the 19th century, took the most detailed measurements of the 
Great Egyptian Pyramids. Having a great set of figures at his disposal and considering 
their different arithmetic combinations, he achieved extraordinary results: with a very 
high level of proximity he, for example, obtained the number $\pi$ , calculated 
the average distance between the Earth and the Moon, and got the parameters of other 
astronomic concepts. Piazzi Smyth came to a definite conclusion: the builders of the 
Pyramids possessed knowledge unknown to the inhabitants of the Earth at that period 
of time, they were extremely technically skilled and consequently were the 
representatives of non-Earth civilization. But the real explanation of the effect is rather 
simple: if one has a complete set of numbers and a certain freedom to manipulate them, 
one can achieve any result wanted.

$$ $$
{\bf 2. 	THE PHENOMENON OF LARGE NUMBERS COINCIDENCE}
{\medskip}

	According to the common view, this phenomenon proves the existence of some 
deep connection between submicro- and mega-physics. Large numbers and the 
phenomenon of their coincidence were first mentioned in H.Weyl's 
works [1,2]. Later, this problem was thoroughly tackled by A.Eddington 
[3-5], and in the 30's of the last century it was P.A.M.Dirack who turned 
to the topic in connection with his hypothesis of the changeability of fundamental 
constants.[6]
	The essence of the phenomenon is simple. 

The relation between the intensity of 
electromagnetic and gravitation interactions of elementary particles (of an electron, for 
example) is a first illustration of a large number.
	$$N_1=\frac{e^2/{\hbar}c}{Gm_e^2/{\hbar}c} \approx 10^{40}. \eqno(2.1)$$

	Another large number appears in a different, metagalactical context as the ratio 
of  "the Universe radius"  (the Hubble radius) $R_{Hb}=c/H_0$   to  the  electron  
radius $  r_e=e^2/m_ec^2 $:
$$N_2=m_ec^3/H_0e^2 \approx 10^{40}, \eqno(2.2)$$
where $H_0$  -  is the Hubble constant.

	As P.Jordan first discovered [7] , the ratio of the mass of a 
typical star $M_*$ to the electron mass is also connected with large 
numbers:
	$$M_*/m_e \approx 10^{60} \approx N_1^{3/2}. \eqno(2.3)$$
	
	Jordan was too emotional about these figures since he thought them to be the 
precursor of cardinal revolutionary changes in cosmology.

	Estimating the average matter density in the Universe $\rho \approx 10^{-30} g/cm^3$ , we 
might  consider the ratio of "the mass  of the Universe" $M_0$ to the proton 
mass which gives the square of the large number:
$$M_U/m_p \approx 10^{80} \approx N_1^2 . \eqno(2.4)$$
	
	Transforming this formula regarding $m_p$ , we might obtain an 
approximate formula which derive mass through the Hubble constant and fundamental 
constants.
	In this context, S.Weinberg [8] gives the empirical formula for the 
pion mass:
$$ m_{\pi}=\left(\frac{\hbar^2H_0}{Gc}\right)^{1/3} - \eqno(2.5) $$
as possessing a real though enigmatic sense. No matter how strange it might 
seem to seriously speak about the formulas which reveal the mass of elementary 
particles through cosmological parameters, such relations, as we will see later, are 
fairly real.
$$ $$
{\bf 3. A NEW APPROACH TO THE LARGE NUMBERS }

{\bf COINCIDENCE}

	Our approach to the phenomenon of large numbers is simple and natural. We will 
discuss the ratio of cosmological parameters to the corresponding microscopic 
parameters: for example, the ratio of the largest parameter of length, the Hubble 
radius, to the smallest one - the Plank length. We will later give a few examples of 
large numbers coincidence. For estimation, we will use the value of the Hubble 
constant $H_0=75 km/Mps \cdot á$. So, the ratio of the Hubble radius 
$R_{Hb}=c/H_0$ to the Plank length $l_P=(\hbar G/c^3)^{1/2}=1.6 \cdot 10^{-33} cm $
 is $$R_{Hb}/l_P \approx 10^{60} . \eqno(3.1) $$
.

	The ratio of the "Universe mass" to the Plank mass produces a figure of the 
same order
$$M_{Hb}/m_P \approx 10^{61}. \eqno(3.2)$$
	From this relation follows for the ratio of the Plank mass density to the observed 
matter density of the Universe
$$ \rho_{P}/\rho \approx 10^{120} . \eqno(3.3) $$

	The square of the ratio of the Plank energy to the background microwave 
radiation temperature $ T_\gamma=2.726 {}^o Š $ is 
	$$(E_P/T_\gamma)^2 \approx 10^{60} . \eqno(3.4)$$
	
	Estimating the neutrino mass $m_\nu \approx 10^{-3}-10^{-4} eV $, we will obtain the 
following for the square of the ratio of the Plank mass to the neutrino mass:
	$$(m_P/m_\nu)^2 \approx 10^{61}. \eqno(3.5) $$
	
	The cube of the ratio of the Plank mass to the mass of elementary particles is 
$$ (m_P/m_e)^3 \approx 10^{62} , (m_P/m_\pi)^3 \approx 10^{63}, (m_P/m_p)^3 \approx 10^{58}. \eqno(3.6)$$

	Let us adduce here the ratio of the typical star mass (the limit of 
Chandrasekhar-Landau) to 
the electron mass
	$$M_*/m_e \approx 10^{60} . \eqno(3.7)$$
	
	The list of coincidences is rather long, as you can see, and this list might be 
prolonged but we will turn to the further discussion of this problem after we have 
analyzed the Hubble and Plank scales.
	
$$ $$
{\bf 4. THE PLANK SCALES}
{\medskip}

	Let us give the value of the fundamental constants which are necessary for the 
future.
$$ \hbar = 1.05457 \cdot 10^{-27} erg\cdot s, \ \ \  G=6.673\cdot10^{-8} sm^3/g\cdot s^2$$
$$ c=2.99792\cdot10^{10} sm/s, \ \ \ \alpha=e^2/\hbar c=1/137.035999. $$
	
	We use two Plank scales. The first Plank scale of quantity is somehow different 
from the traditional one.
$$m_{Pl}=\frac{1}{2}\left(\frac{\hbar c}{G}\right)^{1/2}=1.0884\cdot10^{-5} g, $$
$$ l_{Pl}=2\left(\frac{\hbar G}{c^3}\right)^{1/2}=3.232\cdot10^{-33} cm, $$
$$ t_{Pl}=2\left(\frac{\hbar G}{c^5}\right)^{1/2}=1.078\cdot10^{-43} s, $$
$$ E_{Pl}=\frac{1}{2}\left(\frac{\hbar c^5}{G}\right)^{1/2}=6.11\cdot10^{18} GeV, $$
$$ \omega_{Pl}=1/t_{Pl}=0.928\cdot10^{43} s^{-1} . \eqno(4.1)$$	
	
	The second (reduced) Plank scale differs from the first one by the factor 
$\alpha^{-1/2}$= 11.706237, i.e.
	$$ m_0=\alpha^{1/2}m_{Pl}=0.9298\cdot10^{-6} g,$$
$$r_0=\alpha^{-1/2}l_{Pl}=3.7835\cdot10^{-32} cm, $$
$$t_0=\alpha^{-1/2}t_{Pl}=1.262\cdot10^{-42} s, $$
$$E_0=\alpha^{1/2}E_{Pl}=0.523\cdot10^{18} GeV, $$
$$\omega_0=1/t_0=0.792\cdot10^{42} s^{-1}.  \eqno(4.2)$$

	Besides these two scales, we will need to introduce the mass with the value
$$m_*=2\alpha^{-1/2}m_{Pl}=2.548\cdot10^{-4} g. \eqno(4.3)$$
		
	The main characteristic of the mass $m_*$ is the 
equality of its gravitation radius to the reduced Plank length $r_0$
from (4.2).
	
$$ $$
{\bf 5.THE HUBBLE SCALE AND COSMOLOGICAL}
{\bf PARAMETERS}
{\medskip}

	For the last few years, radical changes in observed cosmology and astrophysics 
have taken place thanks to the realization of the more than 50 projects dealing with 
research of  background microwave radiation (see their description in 
[12]). The results of that research (primarily of the projects COBE, 
BOOMERANG, MAXIMA ) provivded  very important information about 
cosmological parameters and made their value more precise.

	Below, basing oneself  on the analysis of those data 
[10-14], we present the magnitudes of the Hubble scale 
parameters. The Hubble constant in the traditional units is
	$$67<H_0<77 (km/Mps\cdot s) \eqno(5.1)$$
	and in Hertz
$$2.17\cdot10^{-18} s^{-1}< H_0< 2.5\cdot10^{-18} s^{-1}. \eqno(5.2)$$
	
	The time parameter of the scale (the Hubble time) $t_{Hb}=1/H_0$
	$$4\cdot10^{18} s < t_{Hb} < 4.6\cdot10^{18} s. \eqno(5.3)$$
	The parameter of the length (the Hubble radius) is defined as 
$$ 1.2\cdot10^{28} cm <R_{Hb}<1.4\cdot10^{28} cm. \eqno(5.4)$$
	The parameter of the mass (the Hubble mass) is defined as 
$$M_{Hb}=c^3/2H_0G. \eqno(5.5) $$
	The value is within the limits
$$0.81\cdot10^{56} g <M_{Hb}<0.93\cdot10^{56} g.\eqno(5.6)$$
	The Hubble mass density presented as the ratio of $M_{Hb}$ to the 
Hubble volume $V_{Hb}=\frac{4}{3}\pi R_{Hb}^3 $
$$\rho_{Hb}=\frac{3H_0^2}{8\pi G} \eqno(5.7)$$
	coincides with the well-known parameter - the critical density of the matter $\rho_{cr}$.
	
	The value of the critical density:
	$$ 0.843\cdot10^{-29} g/cm^3 <\rho_{cr}< 1.12\cdot10^{-29} g/cm^3. \eqno(5.8)$$ 
	
	The energy density (the mass density), bound to $\Lambda$ , of the 
gravitational field, equation is among other very important cosmological parameters. In 
relative figures -$\Omega_\Lambda=\rho_{\Lambda}/\rho_{cr}$, where the mass density 
$$ \rho_{\Lambda}=\frac{\Lambda c^2}{8\pi G }.$$
		
	The measurements localized $ \Omega_{\Lambda}$ within the limits [16] : 
$$0.5<\Omega_{\Lambda}<0.8. \eqno(5.9)$$
	
	The corresponding value $\Lambda $ is
	$$0.99\cdot10^{-56} cm^{-2}<\Lambda<1.6\cdot10^{-56} cm^{-2}. \eqno(5.10)$$
	
	The next parameter "the dark mass density", $\Omega_{\Delta}$ is usually 
bound to the nonbarionic matter forms, which are considered to be distributed 
throughout the Universe. We will consider the parameter $\Omega_{\Delta}$ in 
analogy with $\Omega_{\Lambda}$, introducing $\Delta$ close 
to $\Lambda$. Thus, 
		$$\rho_{\Delta}=\frac{\Delta c^2}{8 \pi G}. \eqno(5.11)$$
	
	The value $\Omega_{\Delta}$ is localized  within
	$$0.25<\Omega_{\Delta}<0.45.\eqno(5.12)$$

	Correspondingly the value of the parameter $\Delta$:
$$0.5\cdot10^{-56} cm^{-2}< \Delta < 0.9\cdot10^{-56} cm^{-2}.\eqno(5.13)$$
	
	Mutual densities $\Omega_{\Lambda}+\Omega_{\Delta} \approx 1$ significantly exceed the density of 
the usual barionic matter $\Omega_b$, the value of which is 
$$0.03< \Omega_b<0.06.\eqno(5.14)$$
	
	In conclusion, we  present the value of the cosmological parameter, whose 
measurements are known best of all -  background microwave radiation temperature
$$T_{\gamma}=2.726 {}^o K \approx 2.349\cdot10^{-4} eV. \eqno(5.15) $$
$$ $$  
$$ $$  
{\bf 6. THE HUBBLE CONSTANT : ONE CAN HARDLY }

{\bf BELIEVE IN IT}
{\smallskip}

	Let us discuss the ratio of the Hubble and Plank values in a 
more detailed  and  thorough  way.  We  take  the  Plank  value from the  reduced scale
	(4.2). It is obvious $R_{Hb}/r_0=t_{Hb}/\tau_0=\omega_0/H_0$. 
For the last ratio of the Plank frequency to the Hubble constant we have 
	$$3.17\cdot10^{59}<\omega_0/H_0<3.65\cdot10^{59}.$$
	The natural logarithm of the ratio
$$137.006<\log (\omega_0/H_0)<137.147.\eqno(6.1)$$
	
	It is extremely surprising! The inverse value of the fine structure constant 
$1/\alpha=137.035999$ can fit this narrow interval! What is it? An extraordinary coincidence or 
fact which has some deep physical sense? Taking into consideration the numerous 
examples of the large numbers coincidence (3.1) - (3.7), we tend towards the 
favouring of the latter and we make the following 

	{\bf SUPPOSITION A.}

{\it	The logarithm of the ratio of the Plank frequency $\omega_0$ to 
the Hubble constant is equal to the inverse value of the fine structure constant:}
	$$\log(\omega_0/H_0)=1/\alpha.\eqno(6.2)$$
	
	From this immediately follows
$$H_0=\omega_0 e^{-1/\alpha} ,\eqno(6.3)$$
	which looks to be too fascinating  in its full form
$$H_0=\frac{ec^2}{2\hbar \sqrt{G}}e^{-\hbar c/e^2}.\eqno(6.4) $$
	
	All this is definitely very strange and not quite understandable. Why should the 
Hubble constant which characterizes the speed of the Universe expanding  be 
connected with the fundamental constants?
	By analogy, we have
$$R_{Hb}=r_0e^{1/\alpha}, \ \ \  t_{Hb}=\tau_0e^{1/\alpha} .\eqno(6.5) $$
	
	We can also write down the value of the Large Number ${\cal B}_0=e^{1/\alpha}$:
			$${\cal B}_0=0.326572\cdot10^{60} \eqno(6.6)$$
	and for the reference some values
$${\cal B}_0^{-1}=3.062115\cdot10^{-60}, $$
$$ {\cal B}_0^{1/2}=0.57146\cdot10^{30}, $$
$${\cal B}_0^{-1/2}=1.74989\cdot10^{-30},$$
$$ {\cal B}_0^{1/3}=0.688641\cdot10^{20}, $$
$${\cal B}_0^{-1/3}=1.452136\cdot10^{-20}. \eqno(6.7)$$

	From the formula (6.3), we can define the "exact" value of the Hubble constant:
	$$H_0=2.425\cdot10^{-18} c^{-1} = 74.85 km/Mps\cdot s .\eqno(6.8)$$
	
	If we can not be absolutely sure of the exactness of the formula (6.4), then at 
least, we will not have any doubts as to the fact that (6.4) gives the main value of the 
Hubble constant and the only problem is in the possibility of slight  corrections. Thus, 
strictly speaking, the inequality might be written as the more general expression than (6.2)
$$\log(\omega_0/H_0)=\frac{1}{\alpha}+ O(\alpha) .\eqno(6.9)$$
	
	In this connection, we will make the following

	{\bf SUPPOSITION A'}

{\it	The  logarithm of the ratio of the Plank frequency $\omega_0$ to the 
Hubble constant equals the inverse value of the fine structure constant with slight 
corrections $\alpha$}
	$$\log (\omega_0/H_0)=\frac{1}{\alpha}+a_1\alpha + a_2\alpha^2+ \dots \ \ \ .\eqno(6.10)$$

	The more general supposition introduces some uncertainty  in the value 
$H_0$. But this uncertainty is not very big and is less than one per 
cent.
	The following numerological experiment is a kind of proof and an illustration to 
the pertinence of the specifying suppposition made above. For the experiment, we will 
use the parameter $T_\gamma$, suppposing that it is known 
approximately
	$$2.7 K<T_{\gamma}<2.75 K \eqno(6.11)$$ 
		and we will try to numerologically reconstruct its exact value. Let us form the 
expression
	$$A=9{\cal B}_0^{-1}(E_0/T_{\gamma})^2 \eqno(6.12)$$
	and define the interval of its localization.
$$134.6<A<138.3. $$
	
	Hence, supposing that $A=1/\alpha$, we will obtain
	$$T_{\gamma}=3\alpha ^{1/2} E_0 e^{-1/2\alpha} \approx 2.722 {}^0 K. $$
	
	It is very close to but not the exact value. Now, in conformity with Supposition 
A' we will present
$$A=\frac{1}{\alpha}+a_1 \alpha+ O(\alpha ^2).$$
	Then
$$T_{\gamma}=3\alpha ^{1/2} E_0 (1+a_1\alpha )e^{-1/2\alpha},  $$
	
	where the unknown coefficient $a_1$ is defined by 
comparison with the exact value $T_\gamma=2.726$. And then
	$$T_{\gamma}=3\alpha ^{1/2} E_0 (1+\frac{1}{5}\alpha )e^{-1/2\alpha}.$$
$$ $$
{\bf 7. PRECISE DEFINITION OF THE PARAMETERS }

{\bf $\Omega_{\Lambda}$ AND $\Omega_{\Delta}$}
{\medskip}

We proceed from the estimations
$$0.5<\Omega_{\Lambda}<0.8 \eqno(7.1)$$
and
$$0.25<\Omega_{\Delta}<0.45 .\eqno(7.2)$$

For the parameters $\Lambda$ and $\Delta$ we 
correspondingly have 
$$0.99\cdot10^{-56} cm^{-2}<\Lambda<1.6\cdot10^{-56} cm^{-2} $$
and
$$0.5\cdot10^{-56} cm^{-2}< \Delta < 0.9\cdot10^{-56} cm^{-2}.$$
Let us construct the expression  $(2/\Lambda r_0^2)^{1/2}$ and  find the 
localization of its logarithm.
$$136.96 < \log (2/\Lambda r_0^2)^{1/2}< 137.20 \eqno(7.3)$$
and  make the next

{\bf SUPPOSITION B}

{\it The logarithm $(2/\Lambda r_0^2)^{1/2}$ is equal to the inverse value of fine 
structure constant.}
$$\log (2/\Lambda r_0^2)^{1/2}=1/\alpha.$$
From this it  can be concluded that
$$\Lambda=\frac{2}{r_0^2} e^{-2/\alpha} $$
or
$$\Lambda=\frac{2}{R_{Hb}^2}. $$
That gives for  $\Lambda$-energy density
$$\Omega_{\Lambda}=2/3 .\eqno(7.4)$$
Having repeated the procedure the same way for 
$\Omega_{\Delta}$ and $\Delta$,  one can get
$$ 136.9< \log \left(\frac{1}{\Delta r_0^2}\right)^{1/2} < 137.21.$$
hence,  making 

{\bf SUPPOSITION C}
{\it The logarithm  $(1/{\Delta} r_0^2)^{1/2}$ is equal to the inverse value of 
the fine structure constant}
$$\log (1/\Delta r_0^2)^{1/2} = 1/\alpha. $$
Hence
$$\Delta=\frac{1}{r_0^2} e^{-2/\alpha}=1/R_{Hb}^2.$$
Accordingly  relative energy (mass) density $\Lambda$
$$\Omega_\Delta=1/3 .\eqno(7.5) $$
Traditionally, the density $\Omega_\Delta$ is connected with the  large quantity 
of  non-barionic matter existing in the Universe  in the form  of weakly interacting 
mass particles (WIMP),  supersymmetrical partners of various particles etc. This non-
barionic matter is regarded as "responsible" for  "dark mass" concentrated in galaxies 
and their groups. But our numerical results, though,  make the aforementioned sound 
doubtful. One cannot  but  feel     that $\Lambda$ and $\Delta $ matters are related. They 
might be regarded as the two sides of the coin. For example, $\Omega_\Delta$ is two 
times smaller than $\Omega_\Lambda$.  If one sum them up , he will get  a one, i.e. 
$\rho_\Lambda+\rho_\Delta=\rho_{cr}$.    The $\Lambda$-matter has  the equation of  state
$$P_\Lambda=-\varepsilon_\Lambda,\eqno(7.6)$$
where $P_\Lambda$ is the pressure, $\varepsilon_\Lambda$ is the density of energy.  

Concerning the equation of the $\Delta $-matter state and status equation the 
following radical 

{\bf SUPPOSITION D}

can be made:

{\it $\Delta$-matter as well as $\Lambda$-matter  is  of  exotic nature and can 
be described with the help of  the equation of state}

$$P_\Delta=\varepsilon_\Delta, \eqno(7.7)$$
where $P_\Delta$ is pressure, $\varepsilon_\Delta=\rho_\Delta c^2$ is the density of energy.

Considering $\Delta$ and $\Lambda $ matteries alltogeather, 
one can  write down the combined equation of  state:
$$P_\Delta+P_\Lambda = \varepsilon_\Delta - \varepsilon_\Lambda $$
or
$$P=- \frac{1}{3}\varepsilon ,$$
where
$$ P=P_\Lambda + P_\Delta , \eqno(7.8)$$
$$ \varepsilon = \varepsilon_\Delta + \varepsilon_\Lambda .$$

If one considers  $\Lambda$ and $\Delta$ matteries as the two 
sides of the coin,   the latter, in this context,  might be regarded  as some united exotic 
environment, which will be further called the "cosmological vacuum" (C-vacuum, 
quintessence). Let's think of the evolution of the Universe  in which   there is no 
matter of any kind and which is filled  only with  the C-vacuum. Within the framework 
of  the Standard cosmological model  the equations of  the gravitation  field 
look like the following ( Fridman equations):  
$$\frac {1}{2} \left( \frac{da}{dt} \right)^2=\frac{4\pi G}{3c^2}a^2(\varepsilon_\Lambda + \varepsilon_\Delta), \eqno (7.9) $$
$$\frac{d^2a}{dt^2} =- \frac {4\pi G}{3c^2} (\varepsilon_\Lambda + 3 P_\Lambda  + \varepsilon_\Delta + 3 P_\Delta) \eqno(7.10) $$

As the C-vacuum density is equal to the critical one it can be concluded that the 
Universe is flat. And  according to the equations of  state (7.6), (7.7), and 
$\varepsilon_\Delta=\varepsilon_\Lambda /2$ the right-hand  expression in brackets (7.10)  becomes 
equal to zero, and instead of (7.9) and (7.10) we have
$$ {\dot a}/{\ a}=H_0 , $$
$$ {\ddot a}/a=0 , $$

i.e.,  as it should be expected,  the "empty"  flat Universe  expands uniformly and 
with  steady speed on the  condition that  real substance is absent. 

The C-vacuum can be perfectly described in terms of  a real  scalar field. Let us 
briefly consider this point  as well:

Scalar field Lagrangiane looks  as follows [15]:
$$ {\cal L}=\frac{1}{2} \partial_\mu \phi \partial^\mu \phi - V(\phi) , $$

Where the first term of the equation is  the density  of the field kinetic energy, 
the second term is  the density of  the field potential energy.

Stress-energy tensor of the scalar field is
$$ T^{\mu \nu}=\partial^\mu\phi\partial^\nu\phi - g^{\mu\nu}{\cal L} \eqno (7.11)$$

Within the framework of the Standard cosmological model a supposition can be 
made where the components of the tensor are regarded in a perfect fluid 
approximation, for which in terms of energy density and pressure
$$T^{\mu\nu}=(\varepsilon +P)u^{\mu}u^{\nu}-Pg^{\mu\nu}\eqno(7.12) $$

Therefore,
$$T^{00}=\varepsilon ,\ \ \ T^{11}=T^{22}=T^{33}=-P\eqno(7.13)$$

Taking into consideration the cosmological principal,  the scalar field is regarded 
as being homogenous, i.e.  gradients $\nabla \phi$ are equal to zero.   Owing to 
this fact the stress-energy tensor components are:
$$T^{00}=\frac{1}{2}{\dot \phi}^2 + V(\phi), \ \ T^{11}=(\partial^1 \phi)^2-V(\phi), $$
$$T^{22}=(\partial^2 \phi)^2-V(\phi), \ \ T^{33}=(\partial^3 \phi)^2-V(\phi). $$
Hence, the pressure is
$$P=\frac{1}{3}(T^{11}+T^{22}+T^{33})=\frac{1}{2} {\dot \phi}^2 -V(\phi) \eqno(7.14)$$                                                               
and energy density is                
$$\varepsilon = T^{00}=\frac{1}{2}{\dot \phi}^2 + V(\phi). \eqno(7.15)$$
This coincides with (7.8)  for the C-vacuum components. Thus one can write
$$\varepsilon_\Delta=\frac{1}{2}{\dot \phi}^2,$$
$$\varepsilon_\Lambda=V(\phi),$$
$$P_\Delta=\frac{1}{2}{\dot \phi}^2,$$
$$P_\Lambda=-V(\phi).$$
i.e. $\varepsilon_\Delta$ is  the density of  the C-vacuum kinetic 
energy, $\varepsilon_\Lambda$ is the density of potential energy.  Without going into details,  
it should be mentioned  in conclusion  that  cosmological vacuum is not in any regard 
to be equated with  that one of quantum field theory,  the density of which  is 
${\cal B}_0^2=10^{119}$ times higher, i.e. $\rho_{QFT}=\rho_{cr}e^{2/\alpha}$  .
$$ $$
{\bf 8.BARIONIC MATTER DENSITY}
{\medskip}

Due to the newest measurements it is known that the barionic matter density is within
$$0.03<\Omega_b<0.06.$$
The barionic mass of the Universe (the mass of  barions inside  Hubble sphere)
$$2.3\cdot10^{59}<M_b<4.6\cdot10^{59}.$$
The logarithm
$$136.7< \log (M_b/m_{Pl})<137.4,$$
One make the aforementioned

{\bf SUPPOSITION  E}

{\it  The logarithm $M_b/m_{Pl}$ is equal to the inverse value of the fine structure 
constant.}
$$\log (M_b/m_{Pl})= \frac{1}{\alpha}+O(\alpha). $$

Following the supposition  and taking into consideration that $M_b=m_{Pl}\cdot{\cal B}_0$
and $M_{Hb}=m_*{\cal B}_0$  a simple analytic expression is available 
for barionic matter density of the Universe
$$\Omega_b=M_b/M_{Hb}=\frac {1}{2}\alpha^{1/2}. $$

Numerical value of  barionic density is
$$\Omega_b \approx 0.047.$$

Total density of the matter of the Universe is
$$\Omega_{tot}=1+\Omega_b+\Omega_\nu+\Omega_\gamma+\Omega_G+\dots.$$

C-vacuum density ($\Omega_\Lambda+\Omega_\Delta=1$), as well as barionic, $\Omega_b=0.047$  predominate in total density. Further, one 
will  think that
$$\Omega_{tot} \approx 1+ \frac{1}{2}\alpha^{1/2}.$$
As the density of   the other forms of matter is to be less than $\Omega_b$,
otherwise it would contradict the observation data. But here one can face  a problem 
of "dark mass" in galaxies and galaxies groups.  Not to go into details  two ways of  
solving the problem  should be mentioned here:

1.	The dark mass effect being observed  in galaxies and theirs groups is a 
mirage. The cause for the mirage is the changing of  the form of the gravitation 
interaction at  great distances 
	($R>>R_0 \approx 10 kps$) .  For example,  gravitation potential 
	$$V(r)=-\frac {GM}{r} + \frac {GM}{R_0} \log \frac {r}{R_0} $$
	may be used to describe the effect in  the proper way.
	
$$ $$
2.	Dark matter is connected with C-vacuum desity increas it the vicinity of 
heavily gravitating objects. 
$$ $$
{\bf 9. THE AGE OF THE UNIVERSE AND THE }
{\bf PARAMETER OF RETARDATION}
{\medskip}

The Age of the Universe 
$$t_U=\frac {b}{H_0} ,\eqno(9.1)$$

Here the parameter b is depend on a concrete type of the 
cosmological model under consideration.
It can be concluded from (9.1) that
$${\dot H_0}=-\frac{1}{b}H_0^2 .\eqno(9.2)$$

According to definition the retardation parameter is $q_0=-{\ddot a}/aH_0^2$.
On the one hand, from  the Hubble formula ${\dot a}=H_0a$ one 
can get
$$q_0=\frac{1}{b}-1,$$
On the other hand, from the second Fridman equation one can get
$$-\frac{{\ddot a}}{aH_0^2}=\frac{4\pi G}{3c^2 H_0^2}\cdot\frac{\rho_b c^2}{2}, $$
or
$$q_0=-\frac{{\ddot a}}{aH_0^2}=\frac{1}{2}\Omega_b. $$

Finally, for the retardation parameter and parameter b
$$q_0=\frac{1}{4}\alpha^{1/2},$$
$$b \approx 1 -\frac{1}{4}\alpha^{1/2}.$$
Thus, the age of the Universe is 
$$t_U=\frac{1-\alpha^{1/2}/4}{H_0} \approx 0.4\cdot10^{18} c \approx 12.8\cdot10^9 \ \ years$$
$$ $$
{\bf 10. CHANGEABILITY OF FUNDAMENTAL CONSTANTS }
{\medskip}

P.A.M.Dirak was the first to suggest the idea of changeability of fundamental 
constants in the context of the large numbers coincidence phenomenon. On the whole 
the idea was perceived with a slight doubt. Nevertheless, the results of this research 
which have revealed the analytic interconnections between fundamental constants and 
the changing-in-time cosmological parameters do   leave no room for the  doubt about 
the changeability of fundamental constants. From the author's point of view, it should 
be noticed that, even without taking into consideration the aforementioned, in the non-
stationary Universe   all  physical quantities, including fundamental constants, are 
bound to change.

It is the speed of fundamental constant changeability  which is of interest. The 
general  rate of the Universe changeability,  the speed of its expanding is determined 
with the the Hubble constant value.
Therefore, the following plausible 

{\bf SUPPOSITION  F}
\ \ can be made:

{\it Fundamental constants F changes in time with the relative 
speed which is  proportional  to the Hubble constant.}

$$\frac{{\dot F}}{F} \sim H_0. $$

How can this supposition be supported? First, it should be noticed (as we can see  
from (9.2)) that  the Hubble constant itself  changes in time with speed which 
is

$$\frac{\dot{H}_0}{H_0}=-\frac{1}{b}H_0 $$

Further, from the expression (6.3), as well as from many others analogous ones  
one can find directly
$$\frac{\dot{ \omega_0}}{\omega_0} = \frac{\dot{H}_0}{H_0} - \frac{{\dot \alpha}}{\alpha^2} \sim H_0$$

which, on the whole, is for the mentioned supposition.
$$ $$
{\bf 11. MASS OF ELEMENTARY PARTICLES}
{\medskip}

To draw the line let us consider the point which, on the one hand, seems to be 
far from  cosmology, 

But on the other hand it deals with  large numbers and  numerology  directly.  As 
it was above  mentioned,  the relation between the Plank mass and  the mass of 
elementary particles has  the large number order

For example, let us consider the mass of an electron.  In order to do this we are 
to evaluate the expression
$$\alpha^{3/2}{\cal B}_0^{-1}\left( \frac{m_0}{m_e} \right) ^3 = 2.0464 \approx 2. $$

Therefore for the electron mass one can get the simple formula
$$m_e=\frac{1}{2^{1/3}}\alpha^{1/2}m_0e^{-1/3\alpha} \approx 0.515 MeV .$$

Or it may be represented in detail
$$ m_e= \frac {1}{2^{4/3}} \frac{e^2}{\sqrt {G\hbar c}} e^{-\hbar c /3e^2}. \eqno(11.1)$$                               

The formula cannot be regarded as a banal approximation, from the author's 
viewpoint.  One would think that one day the formula with correction terms will be 
obtained from first principls.  Within the framework of a some high theory.

Let us consider what we have for the mass of the pion.  The following 
expression is to be calculated by analogy
$$ \alpha^{-1/2}{\cal B}_0^{-1} \left( \frac{m_{Pl}}{m_{\pi ^+}} \right) ^3 \approx 3 .$$
Therefore
$$m_{\pi ^+}=\frac{1}{3^{1/3}} \alpha^{-1/6}m_{Pl}e^{-1/3 \alpha}. \eqno(11.2)$$

Let us use the aforementioned Weinberg empiric formula for the mass of a pion

$$ m_{\pi} \approx \left( \frac{\hbar^2H_0}{Gc} \right) ^{1/3}. $$

After non-complicated transformation we get
$$m_{\pi}=\left( \frac{\hbar c \cdot \hbar \omega_0}{Gc^2}\frac{H_0}{\omega_0}\right) .$$

Taking into consideration that $\hbar c /G = 4\alpha^{-1}m_0^2 ,\ \ \  \hbar \omega_0 = m_0c^2, \ \ \ $
$ H_0/\omega_0={\cal B}_0^{-1}  $,
one can get the formula which is similar to that one  (11.2)
%$$m_{\pi} \approx 2^{2/3} \alpha^{-1\3}m_0e^{-1/3\alpha} .$$
$$m_\pi \approx 2^{2/3} \alpha ^{-1/3} m_0 e^{-1/3 \alpha} $$
$$ $$
{\bf 12. CONCLUSIONS}
{\medskip}
 
So, unusual results have been achieved out of practically nothing.  It is evident 
that numerical analysis made by the author has revealed the elements of  a very deep 
interconnection between micro- and megauniverse.  One would hope that  the revealed 
phenomenon encourages  investigations and the creation  of  proper models.  The 
author does not comment on the results obtained.  But it does not mean, though, that  
he has no ideas concerning them. One should notice, without going into details,  that   
further work on creating adequate physical models  is very likely  to be connected with  
the modern Brane world models.

This research has been done with the comprehensive help of the Centre for 
Advanced Research "Theoretical and mathematical physics". The author 
expresses his gratitude to Smurniy Ye.D. for the help with the 
article to be published and to Kutuzova T.S. for  her creative comments.

\end{document}